\documentclass[conference,letterpaper]{IEEEtran}
\usepackage{fancyhdr,graphicx}
\usepackage{fancyhdr}
\begin{document}

\title{A Simple Mechanism for Focused Web-harvesting}

\author{\authorblockN{Z. Akbar and L.T. Handoko}
\authorblockA{Group for Theoretical and Computational Physics, Reearch Center for Physics, Indonesian Institute of Sciences (LIPI)\\
Kompleks Puspiptek Serpong, Tangerang 15310, Indonesia\\
zaenal@teori.fisika.lipi.go.id}}

\maketitle

\begin{abstract}
The focused web-harvesting  is deployed to realize an automated and comprehensive index databases as an alternative way for virtual topical data integration. The web-harvesting has been implemented and extended by not only specifying the targeted URLs, but also predefining human-edited harvesting parameters to improve the speed and accuracy. The harvesting parameter set comprises three main components. First, the depth-scale of being harvested final pages containing desired information counted from the first page at the targeted URLs. Secondly, the focus-point number to determine the exact box containing relevant information. Lastly, the combination of keywords to recognize encountered hyperlinks of relevant images or full-texts embedded in those final pages. All parameters are accessible and fully customizable for each target by the administrators of participating institutions over an integrated web interface. A real implementation to the Indonesian Scientific Index which covers all scientific information across Indonesia is also briefly introduced.
\end{abstract}

\IEEEpeerreviewmaketitle

\section{Introduction}

\begin{figure*}[t]
 \centering
 \includegraphics[width=18cm]{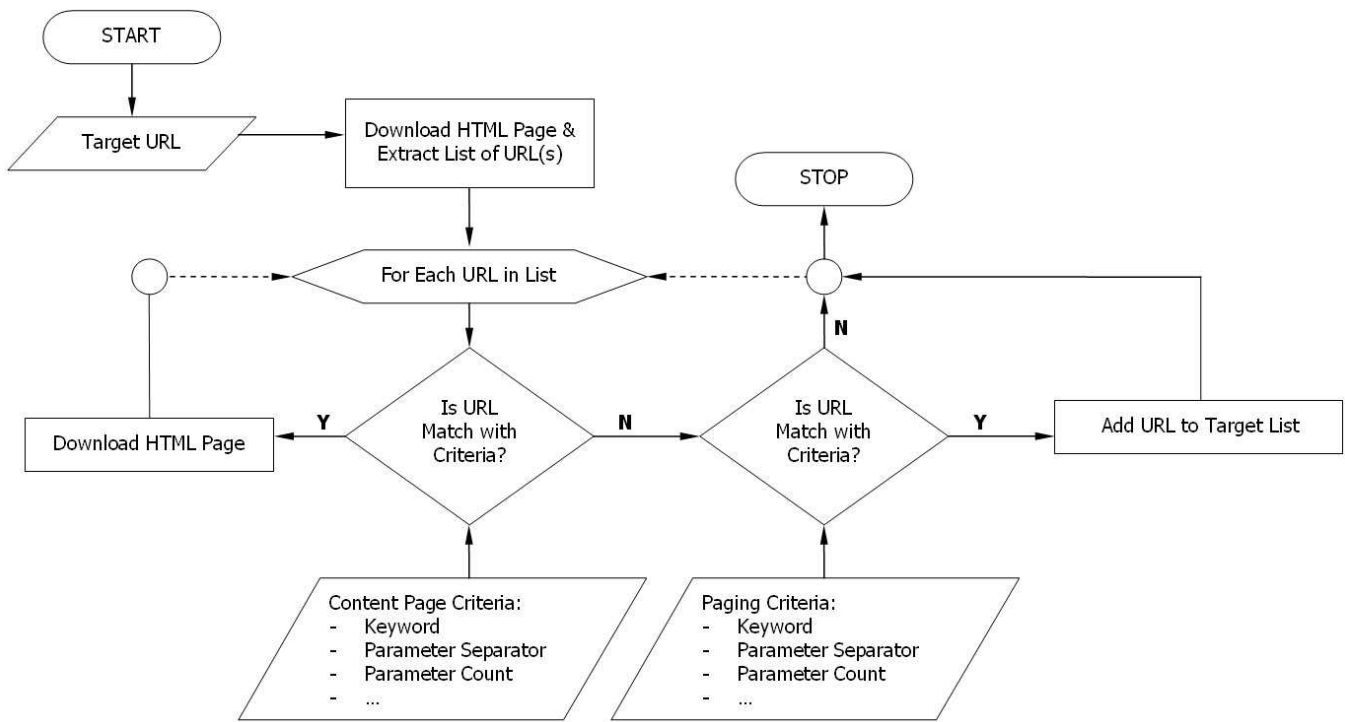}
 \caption{The flowchart of harvesting mechanism.}
 \label{fig:harvesting}
\end{figure*}

Data integration is recently the center issue among the information management communities. Because data integration is intended to overcome the phenomena of information flooding, and on the other the information islands. The second one refers to a condition of separating data pools, though within the same topic, which are not well connected nor integrated each other. Both obscure the potential users to access and to efficiently use the available data. Although data archiving is an important aspect of information and knowledge management since long time ago, it would unfortunately not benefit the  stakeholders without improving the accessibility to the data itself.

There are several methods to establish either real or virtual, and partial or total data integration. Some widely implemented methods can be listed as follows :
\begin{itemize}
\item Conventional method through manual collection and retrieval : \\
This is the typical and old-fashioned method which is in general not supported anymore for a large scale data integration. Although, needless to say that this method guarantees the accuracy and quality of integrated data.
\item Electronic integration over dedicated network :\\
In this system all participating databases remain at their original places, but all of them are connected and integrated at real-time basis through a secure private network. This method is rather costy, relies highly on the reliability of network, requiring a uniform platform and applications among the participating databases. Though, it would keep the accuracy as the conventional method.
\item Conventional search engine : \\
This method is categorized as virtual data integration. Because it integrates the data through the index databases updated in a regular basis. The severe problem is the data retrieval is done through  indiscriminate crawlings of any web pages in relevant sites. It pays the ease with much less accuracy. Moreover, the results often generate another type of information flooding.
\item Federated search : \\
This is recently developed approach to provide a single gateway of search engine enabling simultaneous search at multiple online databases. It is actually an emerging feature of automated,  web-based library and information retrieval systems. However, this requires well connected and online databases. Also the system should be established under official agreements among participating institutions, and requires some modifications at each database to allow query requests from the gateway.
\end{itemize}

Regardless a need for data integration is obvious, in reality there are many non-technical obstacles to realize it. We point out some of them :
\begin{itemize}
\item Real data integration should be realized under a mutual agreement among multi institutions. However establishing such agreements is time consuming and a hard work.
\item Moreover, in that case requirement of modifications or deploying universal standard at each site would increase refusal, since each institution has developed their own system with some uniqueness that might not be able to be accommodated under universal standard. Worsely, there might in some cases be contradictory requirements among them.
\item In most cases, the leading institution is not in a position to force another ones neither to follow certain rules nor to share their data in its original forms. Though, most institutions are eager to advertise and let the public to know their works.
\item Some legal issues, especially property rights, prohibit the data generator to open the data in its original form.
\item Data integration over distributed databases requires numerous number of skilled human resources to maintain.
\end{itemize}
Therefore, no matter how good the idea of data integration is, in most cases it doesn't work as expected. More importantly, the issues are less technical like the data format, etc.

So we should find any intermediate solutions to overcome the problem and to realize data intregation in an efficient manner. For the sake of simplicity, let us focus on the topical data integration. Also by  its nature, the data integration is mostly relevant only for topical integration. In this paper we propose a new method based on the so-called focused web-harvesting. After explaining its concept in the next section, we discuss in detail the general architecture. After introducing its implementation to the Indonesian Scientific Index (ISI), we finish the paper with conclusion and some comments on future developments.

\section{Concepts}

\begin{figure*}
 \centering
 \includegraphics[width=18cm]{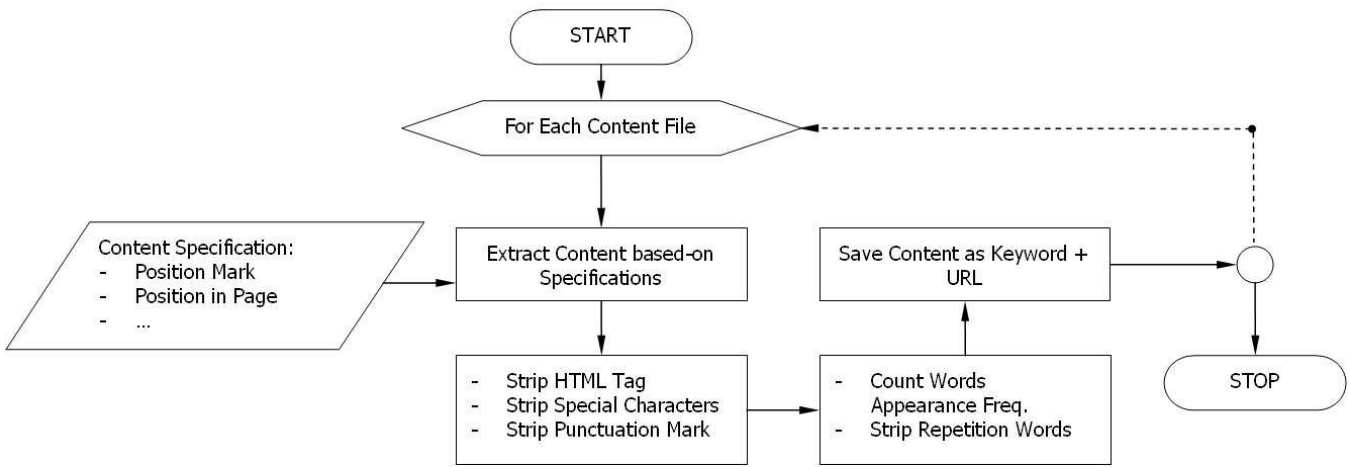}
 \caption{The flowchart of parsing mechanism.}
 \label{fig:parsing}
\end{figure*}

Concerning the backgrounds mentioned in the preceding section, let us assume that,
\begin{itemize}
\item The data integration is limited within a certain topic of information. In the present case we focus on the data relevant to scientific and technology information in Indonesia. 
\item No requirement to the participating institutions, neither modification of their databases nor having sophisticated network infrastructure. This is crucial especially in developing countries where most institutions can not afford sustained commitment for dedicated and stable broadband internet connection.
\item Keep the freedom of each institution to develop their own standards and formats.
\item Instead, there is only a single requirement for all, that is all relevant data should be fully accessible by public over web.
\item No additional burden to participating institutions in any means.
\item Only the indexed data are stored to avoid property right issues. 
\end{itemize}

After some trials in couple of years to established a sustained data integration, we reach at the following integrated solution which can be breakdowned into three components :
\begin{itemize}
\item A centralized infrastructure :\\
There should be a centralized infrastructure hosted and maintained by a leading institution or consortium in the topic. Because once a data integration gateway started providing the service, it would grow very fast and soon requires more financial backup for maintenance and further expansion along with increasing traffics, spaces and memories to handle properly all data.
\item Smart and efficient method to harvest the relevant data : \\
This is the main issue we deal in this paper. The detail is soon given in the next section.
\item Intelligent data analysis to generate accurate index databases : \\
We have deployed text-mining based technology to refine the harvested data and reconstruct well structured and comprehensive index databases for end-users.
\end{itemize}
Actually the first point is consistent with recent facts that successful topical data storages which de-facto integrate all data in some fields are pioneered and hosted in a centralized manner by a leading institution. For example the Astrophysics Data System by SAO \cite{ads}, the preprint repository arXiv pioneered by LANL \cite{arxiv}, the Protein Data Bank by RCSB \cite{pdb} and the DBRiptek by KRT \cite{dbriptek}. Yet, all of them are based on either voluntary or incentive-driven submission by the data owners.

The above integrated solution would benefit all stakeholders, especially the public and participating institutions, that is
\begin{itemize}
\item Assuring up-to-date indexed data due to regular data retrieval as conventional web-crawler.
\item The whole system is relatively low-cost and easy to maintain, since everything is centralized. This is good from limited financial backup and human resources point of view.
\item It also leads to guaranteed durability and sustainability of services in long period.
\item No additional works and certain softwares required in each institution.
\item At least the system can be expected to be more accurate than indiscriminate web-crawling.  Moreover, the accuracy can in principle be improved by implementing advanced text-mining.
\end{itemize}

\section{Architecture}

\begin{figure*}
 \centering
 \includegraphics[width=18cm]{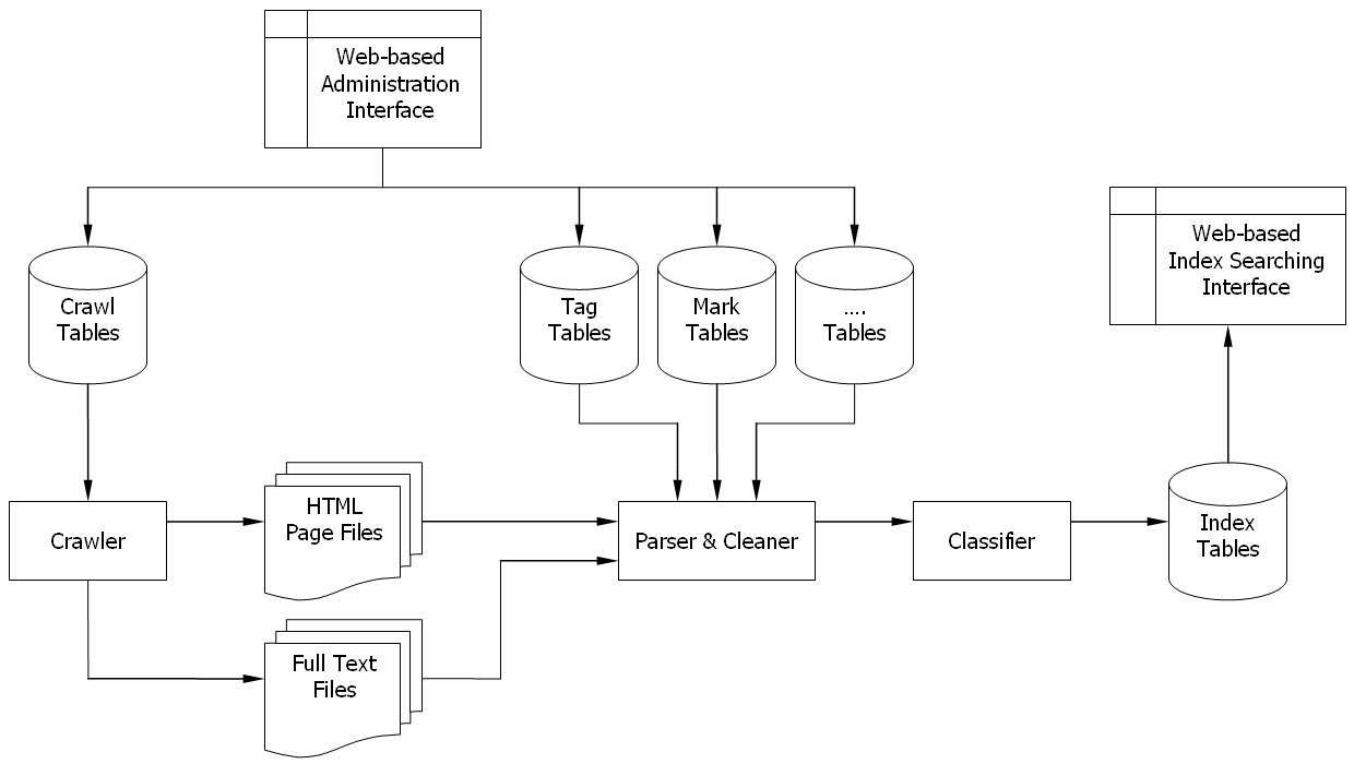}
 \caption{The whole components composing the focused web-harvesting applied to ISI.}
 \label{fig:komponen}
\end{figure*}

Now we are ready to discuss the architecture to realize the mentioned integrated solution. The architecture is inspired and combination of focused web-crawling and web-harvesting. The focused web-crawling does not indiscriminately crawl the web pages like general purpose search engines, but attempts to download pages that are similar to each other  \cite{menczer1}\cite{menczer2}\cite{chakrabarti}. Therefore a focused crawler must predict the probability that a link to a particular page is relevant before actually downloading the page. A possible predictor is the anchor text of links \cite{pinkerton}. However, it clearly should be armed with more sophisticated techniques such as reinforcement learning and evolutionary adaptation to obtain better performance over longer crawls \cite{menczer3}. For instance, one can use    complete content of the pages already visited to infer the similarity between the driving query and the pages that have not been visited yet \cite{diligenti}. The performance of a focused crawler depends mostly on the richness of links in the specific topic being searched, and focused crawling usually relies on a general web search engine for providing starting points.

On the other hand, conventional web-harvesting allows web-based search and retrieval applications such as search engines to index content that is pertinent to the audience for which the harvest is intended. Such content is thus virtually integrated and made searchable as a separate web application. General purpose search engines, such as Google and Yahoo! index all possible links they encounter from the origin of their crawl. In contrast, search engines based on web-harvesting only index URLs to which they are directed. This implementation strategy has the effect of creating a searchable application that is faster, due to the reduced size of the index; and more accurate to generate topical results since the indexed URLs are pre-filtered for the topic or domain of interest. In effect, harvesting makes otherwise isolated islands of information searchable as if they were an integrated whole.

Based on the targeted URLs defined by human expertise or machine guidance, web-harvesting begins to download this list of URLs. Embedded hyperlinks that are encountered can be either followed or ignored, depending on human or machine guidance. A key differentiation between web-harvesting and general purpose web-crawlers is that for web-harvesting, crawl depth will be defined and the crawls need not recursively follow URLs until all links have been exhausted. The rest is the same, that is the downloaded content is  indexed and made searchable over web.

\subsection{Harvesting mechanism}

In order to improve the accuracy and avoid wasting the resources to crawl irrelevant web pages, we have adopted the conventional web-harvesting with more human-guidance parameters setup. The whole mechanism is reflected in the following initial procedure for each target and should be done by the administrators of participating institutions over web :
\begin{enumerate}
\item Determining a specific targeted URL with information relevant to the topic of data integration being built. The URL should direct to the first page of desired type of contents.
\item Counting the depth=scale of crawled page from the initial page directed by the targeted URL to the final page where the relevant contents are harvested.
\item Putting some criterion related to the pages between the initial and final pages to distinguish the embedded hyperlinks direct to a title of final content or page numbers of further title lists. This paging criterion comprises some parameters including keywords, parameter separator, parameter count and so on. 
\item Determining the focus-point number of the desired content in the final page. The number points to the ``box'' containing relevant content. User can define a parameter as the tag for this purpose, for instance \texttt{table} tag from the page top to the box. This marking method is quite effective to zoom in the relevant information. 
\item Putting some keywords to recognize the hyperlinks inside the box. Either the link directs to a relevant image, full-text or anything else. Similar to the paging criterion, the content page criterion also includes  keywords, parameter separator, oarameter count and so on. 
\item Choosing the re-harvesting period according to the updating frequency of each type of content.
\end{enumerate}
The same procedure should be done done for each type of contents maintained by the institutions. 

We should emphasize that this procedure is handed over to the administrator of each institution to keep the parameter set of each targeted URL to be accurate. It also avoids unnecessary delay of knowing  design or any other detail changes at the harvested targets, and provides a freedom for the institution to decide what and how their contents are crawled. 

The flowchart of harvesting mechanism is depicted in Fig. \ref{fig:harvesting}. As shown in the figure, in principle the full human guidance targeted URL can be complemented with machine guidance by adopting text-mining based self-learning system in the harvesting mechanism.

Through the above-mentioned procedure, it is clear that the human-guided parameters would reduce significantly crawling of irrelevant information. Also the mechanism gets rid of some policies commonly concerned in regular or focused web-crawlings like :
\begin{itemize}
\item Selection policy :\\
This policy is not more relevant in our approach, since all targeted URLs are well-defined and automatically already filtered in some sense. In other word all pages are considered important. Also, no need to concern about restricting followed links in crawled pages and how to deal with path-ascending crawling, focused crawling and the deep web.
\item Re-visit policy :\\
This policy is also not a big deal anymore since all types of contents are well-defined and have certain characteristic of regular updatings. More than that, this is determined solely by the institution administrator who knows better the appropriate period for harvesting.
\item Politeness policy :\\
The system is web server friendly since it does not adopt any kind of indiscriminate crawling like general purpose search engines. 
\item URL normalization :\\
By definition all targeted URLs in the current mechanism is manually entered. So there is no chance on crawling the same resource more than once. Of course, complementing it with the automated machine guidance would make the URL normalization to be important.
\end{itemize}

\subsection{Parsing mechanism}

The problem now is how to clean up and refine it to build comprehensive index databases. Due to limited space, let us consider the parsing mechanism with minimum cleaning and refining processes. The parsing mechanism is intended to clean the harvested information from the final box, and refine it to build an efficient index database on it.

In contrast with general purpose web-crawling, we deal in most cases with the databases of topical data. Then, the displayed web pages are generated dynamically using the same HTML template in all pages of certain type of data. This allows us to localized the appropriate box. Various design of web pages can be handled by counting the number of particular tag defined previously, e.g. \texttt{table} tag, till the desired location of box. The number is one of the predefined parameter mentioned previously. 

Further procedure is quite trivial and standard. First, all HTML tags are discarded. After taking appropriate part of relevant text, more extensive cleaning and refining works should be performed. However, this topic is out of coverage of the present paper and could be largely varying according to the purpose of data analysis.

A general parsing mechanism is shown as a flowchart in Fig. \ref{fig:parsing}. The minimum parsing mechanism discussed briefly above is up to the second column of charts.
 
\begin{figure}
 \centering
 \includegraphics[width=9cm]{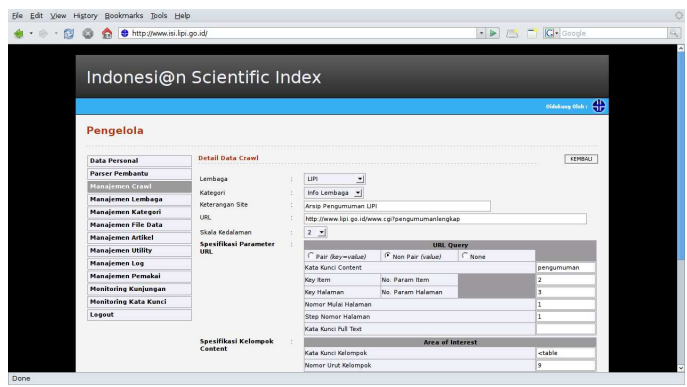}
 \caption{The web interface to predefine a set of harvesting parameters for a certain targeted URL at ISI.}
 \label{fig:isi}
\end{figure}

\section{Implementation : Indonesian Scientific Index}

We have implemented the focused web-harvesting mechanism to build a national scale data integration on scientific and technology information (STI) across Indonesia, namely the Indonesian Scientific Index (ISI). ISI is available online for public since the end of 2007 \cite{isi}. 

So far, ISI covers the databases from few tenth research and academic institutions in Indonesia. We divide the STI into six types : 
\begin{itemize}
 \item Institutional information.
\item Personal (researcher, academician).
\item Scientific publication.
\item Scientific activity.
\item Scientific news.
\item Intellectual property right (paten, copyright, etc).
\end{itemize}
The total targeted URLs for all types of contents reaches more than a hundred with few tenth thousands indexed pages. During the first beta running till March 2008, the algorithms performs perfectly as expected. This might be due to full human-guided parameters setup through the web interface as seen in Fig. \ref{fig:isi}. We have yet not complemented with the automated machine guidance using self-learning systems. 

\section{Conclusion}

We have introduced the concept of focused web-harvesting and its implementation to ISI. The focused web-harvesting differs with the conventional web-harvesting on its wider scale of human interventions, from determining the targeted URLs till setting up all detailed parameters and keywords. 

According to the results from its implementation at ISI, it has shown excellent results as the initial expectation. Although the approach requires more human resources than the general purpose web-crawling, the effort is reasonable to obtain much more accurate data. We argue that the focused web-harvesting is suitable for particular purposes like virtual data integration in certain topics. In contrast, this kind of purposes can not be supported by another web-crawling methods so far.

Now, we are going to incorporate more advanced refining processes using text-mining technology to rebuild well-structured index databases. This is actually a challenging problem concerning so many types of web pages existing around the world. To improve the whole processing speed we also plan to perform parallel web-harvesting. 

Finally, the softwares developed for ISI, its focused web-harvesting and related tools will be open for public under GNU Public License soon after the stable version is completed \cite{openisi}.

\section*{Acknowledgment}

The work is financially supported by the Riset Kompetitif LIPI in fiscal year 2008.

\end{document}